\documentstyle[12pt]{article}
\def\newpage{\vfill\eject}
\font\bigfont=cmr10 scaled\magstep3

\newcommand{\be}{\begin{equation}}
\newcommand{\ee}{\end{equation}}
\newcommand{\mpl}{ {M_{\rm pl}}} 
 
\newcommand\pp{\parshape 2 0.0truecm 15.5truecm 1.25truecm 14.25truecm}
\begin{document}
\baselineskip=22pt
\hsize=15.0truecm
\centerline{}
\bigskip 
\centerline{\bigfont THE LIFE AND TIMES OF EXTREMAL BLACK HOLES} 
\bigskip 
\bigskip 
\bigskip 
\bigskip 
\centerline{\bf Fred C. Adams} 
\medskip 
\medskip 
\centerline{Physics Department, University of Michigan, 
Ann Arbor, MI 48109, USA}
\centerline{fca@umich.edu} 
\bigskip 
\bigskip 
\bigskip 
\centerline{$\Bigl\{$submitted to 
{\it General Relativity and Gravitation}$\Bigr\}$} 
\bigskip 
\bigskip 
\bigskip 
\centerline{\it 15 May 2000} 
\bigskip 
\bigskip 
\bigskip 
\centerline{suggested running head: Life of Extremal Black Holes}

\newpage 
\centerline{\bf THE LIFE AND TIMES OF EXTREMAL BLACK HOLES} 
\bigskip 
\centerline{\bf Fred C. Adams} 
\medskip 
\centerline{Physics Department, University of Michigan, 
Ann Arbor, MI 48109, USA}
\centerline{fca@umich.edu} 
\bigskip 
\bigskip 
\centerline{\bf Abstract} 
\medskip 

{\sl Charged extremal black holes cannot fully evaporate through the
Hawking effect and are thus long lived.  Over their lifetimes, these
black holes take part in a variety of astrophysical processes,
including many that lead to their eventual destruction. This paper 
explores the various events that shape the life of extremal black
holes and calculates the corresponding time scales. }

\bigskip 
\bigskip 
\bigskip 


\bigskip 
\bigskip 
\bigskip 

Extremal black holes contain enough charge so that their electrostatic
energy compensates for their self-gravity. Because they cannot emit
Hawking radiation [1] and do not evaporate, these exotic objects are
often considered to live forever. Extremal black holes do not live in
complete isolation, but rather inhabit a universe destined for eternal
expansion. Because eternity is such a long time, we explore a
collection of astrophysical processes that affect the evolution of
extremal black holes and enforce their ultimate demise.  Many of these
processes take longer than the current age of the universe to operate
and won't become urgent for quite some time.

Stellar and supermassive black holes display substantial astrophysical
evidence for their existence [2]. They are thought to form through
stellar death by supernova or through galaxy formation, respectively.
Although we have no direct evidence for the existence of microscopic
extremal black holes, it nonetheless remains possible for such objects
to be forged in the early universe. Their formation time is expected
to be comparable to the Planck time $\sim 10^{-43}$ sec [3].  These
black holes can have either a magnetic charge or an electric charge,
although we consider only the latter.  (Microscopic black holes
without charge would evaporate long before the present epoch.)
Extremal black holes also provide an important theoretical laboratory
for the study of quantum gravity (e.g., the entropy of a class of
extremal black holes has been calculated from string theory [4]).
Here, we consider possible evolutionary scenarios for extremal black
holes and especially their ultimate fate.

For microscopic extremal black holes, the charge $Q$ required to make
the horizon imaginary is $Q$ = $M/\mpl$, where $M$ is the mass. For
simplicity, we consider the charge $Q$ to be an integer multiple of
the unit electron charge $e$, so that $Q=Ze$. The black hole charge
$Q$ and hence the integer $Z$ can be either positive or negative. The
masses under consideration are thus of order the Planck mass $\mpl$.

The most important processes bearing upon the evolution of extremal
black holes are those that lower their charge through interactions
with particles carrying charges of the opposite sign.  If the black
holes achieve charge neutrality, they rapidly evaporate through the
Hawking effect over a Planck time. 

In the early universe, extremal black holes must directly accrete
particles to alter their charge.  With an effective cross section
comparable in size to the event horizon, $\sigma$ $\sim$ $\mpl^{-2}$,
most interactions occur at the earliest cosmological epochs when the
densities are greatest. Once extremal black holes survive the high
energy environment required for their formation, direct accretion (and
subsequent evaporation) is unlikely.  These exotic objects are thus
likely to survive until the present day.

When the cosmos is $\sim10$ sec old, at the epoch of e$^{\pm}$
annihilation, extremal black holes drop out of kinetic equilibrium and
their internal velocity dispersion falls to $\sim1$ cm/s. Some time
later at $t \approx 10^4$ yr, astrophysical structures start to form.
The universe is thought to contain a substantial admixture of cold
dark matter, weakly interacting particles with a mass density
contribution $\Omega_{CDM} \approx 0.3$.  Both dark matter and
extremal black holes decouple from the background radiation field much
earlier than baryons and begin to collapse before recombination (when
baryonic matter collapses).  The dark matter collects into
self-gravitating structures that eventually become galactic halos and
galaxy clusters. Extremal black holes fall into the deep gravitational
potential wells carved out by the dark matter.  When incorporated into
galactic halos, extremal black holes exhibit dynamical behavior
similar to that of the dark matter and acquire typical velocities 
$v/c \sim 10^{-3}$. 

Once gravitationally confined to a galactic halo, extremal black holes
orbit many times before suffering further interactions. Two important 
processes affect their long term fate: [A] Black holes with positive
charge capture electrons and form bound atomic structures; similarly,
black holes with negative charge interact with protons.  [B] Extremal
black holes pass through stars and stellar remnants, where they are
captured and eventually destroyed.

The galactic disk contains an ample supply of interstellar gas that
can be captured by extremal black holes.  As a reference point, the
recombination cross section for hydrogen is $\sigma \sim 10^{-20} -
10^{-21}$ cm$^2$ under interstellar conditions [5]. With this cross
section and typical number density $n_H \sim 1$ cm$^{-3}$, the
interaction time scale $\tau$ = $1/n_H \sigma v$ $\sim$ $10^6$ yr.
Extremal black holes thus have a reasonably good chance of capturing
charged particles on their passage through the galactic disk. The
limiting factor is the time they spend in the inner portion of the
galaxy (where the gas resides) as opposed to the far reaches of the
galactic halo. Because the gas supply of the galactic disk is expected
to last for $10^{13} - 10^{14}$ yr [6], extremal black holes continue
to make atomic structures over this span of time.

For extremal black holes ($Z=+1$) that successfully capture electrons
and form bound hydrogenic atoms, we can estimate their expected
lifetime.  For a positively charged black hole, the wavefunction of
the electron is similar to that of the hydrogen atom. The ground state
wavefunction is thus $\psi_{100}$ = $(\pi a)^{-3/2} \exp[-r/a]$, where
$a$ is the Bohr radius $a$ = $\hbar^2/m_e e^2$. For the ground state,
the probability ${\cal P}$ that the electron lies within the event
horizon of the black hole is given by 
$$
{\cal P} = 4\pi \int_0^{R_{\rm bh}} | \psi_{100} |^2 \, r^2 dr \, 
\approx {4 \over 3} \bigl( {R_{\rm bh} \over a } \bigr)^3 \, \approx 
{4 \over 3} \bigl( {2 \alpha m_e \over \mpl} \bigr)^3 \sim 
3 \times 10^{-73} \, . 
$$
Folding in the natural oscillation scale of the atom, $t_0 \sim 6
\times 10^{-17}$ sec, we find an atomic lifetime $\tau \sim 10^{49}$
yr. This time scale is much longer than the proton decay time for GUT
processes ($10^{30} - 10^{40}$ yr [7]), somewhat longer than the
proton decay time for gravitational processes ($10^{45}$ yr [8]), and
much shorter than the evaporation time for larger astrophysical black
holes ($t_{\rm evap} \approx$ $10^{65}$ yr $(M_{\rm bh} / M_\odot)^3$
[9]).

For an atomic structure containing a proton orbiting a negatively
charged black hole ($Z=-1$), the Bohr radius is $m_P / m_e$ $\sim$
1800 times smaller.  The proton is an extended particle and the
probability that the black hole lies within the proton is ${\cal P}_1$
$\approx$ $(r_P/a)^3$ $\sim 6 \times 10^{-5}$. The black hole must
accrete one of the proton's quarks in order to change its structure;
this probability is ${\cal P}_2$ $\approx$ $(r_{\rm bh} / r_P)^3$
$\sim$ $10^{-60}$.  The combined probability that the black holes lies
inside the proton and accretes a quark is thus ${\cal P}$ = 
${\cal P}_1 {\cal P}_2$ $\sim$ $6 \times 10^{-65}$. Combining this
result with the oscillation time of the ``atom'', $t_0 \sim 3 \times
10^{-20}$ sec, we find an expected lifetime of $\tau \sim 10^{37}$
yr. On the smallest scale, the net result of this process is a
reaction of the form: q$^{+2/3}$ + bh$^{-1}$ $\to$ q$^{-1/3}$ +
$\gamma$.  Viewed from a larger scale, we see, e.g., p + bh$^{-1}$
$\to$ p e$^{-1}$ $\bar{\nu}$ $\gamma$.

Because the interaction cross section is low, any particular extremal
black hole has a negligible chance of encountering a star during the
current age of the universe ($\sim 10^{10}$ yr).  The galaxy endures
much longer, however, and extremal black holes can eventually
interact. The stars will have long since burned out by the time black
holes pass through them, so the stars are actually stellar remnants --
primarily white dwarfs -- for most of this time [6]. The rate at which
a given extremal black hole passes through stellar remnants is given
by $\Gamma = n_\ast \sigma_\ast v$, where $n_\ast$ $\sim 1$ pc$^{-3}$
is the number density of stars and $\sigma_\ast$ is their
corresponding cross section. Including the effects of gravitational
focusing, the cross section $\sigma_\ast \approx \pi R_\ast^2$
$(1 + 2GM_\ast/R_\ast v^2)$ $\approx$ $3 \times 10^{20}$ cm$^2$ for 
a white dwarf.  With its typical speed $v/c \sim 10^{-3}$, an extremal
black hole encounters a white dwarf every $10^{20}$ years. This time
scale is comparable to the expected galactic lifetime [6,10], the time
required for the galaxy to dynamically relax and evaporate its stars
into intergalactic space. Every extremal black hole should thus
encounter a white dwarf about {\it once} during the lifetime of the
galaxy.

When an extremal black hole enters a white dwarf, the binding
efficiency depends on the rate at which it loses energy as it plunges
through the star. This process is roughly similar to the more well
studied problem of stars collecting large magnetic monopoles. In that
case, main-sequence stars efficiently capture monopoles lighter than
$\sim10^{18}$ GeV and neutron stars efficiently capture all monopoles
lighter than $\sim10^{20}$ GeV [11]. These results imply that the
binding efficiency for charged black holes (with mass $\sim 10^{19}$
GeV) is close to unity for white dwarfs.

After an extremal black hole is confined to a white dwarf, it sinks 
to the center where the particle density is about $n \sim 10^{30}$ 
cm$^{-3}$ for typical remnants. The probability that an electron 
lies within the event horizon of a positively charged black hole is 
${\cal P} \sim 4 \times 10^{-69}$. The oscillation time for degenerate
electrons at this density is about $10^{-20}$ sec, so the time scale
for electron accretion is $\tau \sim 10^{41}$ yr. (Negatively charged
extremal black holes are captured with similar frequency and then
interact with protons in analogous fashion.)

Over vastly longer time scales, any remaining extremal black holes can
interact with electrons or positrons and form immensely large atomic
structures.  The time scale for electrons and positrons to form
positronium in the far future of a flat universe is about $10^{85}$ yr
[12].  The time required for extremal black holes to acquire either
electrons or positrons is thus comparable.  In an open or accelerating
universe, the formation of such atomic structures is very highly
suppressed.  When such atomic structures are created, they are
generally born in highly excited states with extremely large principle
quantum numbers. The time required for these atoms to emit radiation
and spiral down to their ground states is $\sim 10^{141}$ yr [12].
This time is so long compared to the decay time of the ground state
($\sim10^{49}$ yr) that the subsequent annihilation is instantaneous
by comparison. 

The time scales for astrophysical processes that affect extremal black
holes are summarized in Table 1, which also lists times for proton
decay and black hole evaporation. Rather than living forever in stark
isolation, extremal black holes experience a rich and engaging life.
Charged black holes can be created in the very early universe
($10^{-43}$ sec). Their interactions are largely insignificant until
they are incorporated into galactic halos ($10^4 - 10^9$ yr). Once
confined to a galaxy, extremal black holes capture charged particles
and make atomic structures ($10^6 - 10^{14}$ yr). In time, the black
holes accrete their charged partners and radiate away ($10^{49}$ yr).
Extremal black holes are also captured by white dwarfs ($10^{20}$ yr),
where they accrete charge and evaporate ($10^{41}$ yr).  In a flat
universe, extremal black holes that escape destruction by these means
can forge gigantic atomic structures ($10^{85}$ yr), which spiral down
to their ground states and eventually decay ($10^{141}$ yr).  This
timeline presents a rough picture for the life and relevant time
scales of extremal black holes.

\newpage 

\bigskip 
\centerline{}
\bigskip 

\centerline{\bf Table 1} 
\medskip 
\centerline{\bf Time Scales for Extremal Black Holes} 
\bigskip 
 
\begin{center}
\begin{tabular}{lr}
\hline 
\hline
\hline 
Event & Time Scale \\ 
\hline 
\hline 
formation of extremal black holes & $10^{-43}$ sec \\
end of kinetic equilibrium (e$^\pm$ annihilation) $v \to$ 1 cm/s & 10 sec \\ 
collapse into galactic halos begins & $10^4$ yr \\
electron capture in dense gas (fastest rate) & $10^6$ yr\\ 
galactic halos established, $v/c \to 10^{-3}$ & $10^9$ yr \\ 
\hline 
{\it current age of the universe} & $10^{10}$ yr \\ 
\hline 
gas supply depleted, atomic formation ends & $10^{14}$ yr \\ 
accretion by white dwarfs & $10^{20}$ yr \\
{\it GUT scale proton decay processes} & $10^{30} - 10^{40}$ yr \\
ground state atomic decay (bh$^-$ p$^+$) & $10^{37}$ yr \\
destruction within a white dwarf (bh$^+$ e$^-$) & $10^{41}$ yr \\ 
{\it gravitational proton decay processes} & $10^{45}$ yr \\
ground state atomic decay (bh$^+$ e$^-$) & $10^{49}$ yr \\
{\it stellar $(10 M_\odot)$ black holes evaporate} & $10^{68}$ yr \\ 
{\it million solar mass black holes evaporate} & $10^{83}$ yr \\
diffuse atomic structures form in flat universe & $10^{85}$ yr \\ 
{\it billion solar mass black holes evaporate} & $10^{92}$ yr \\
diffuse atomic structures decay in flat universe & $10^{141}$ yr\\ 
\hline 
\hline 
\hline 
\end{tabular}
\end{center}

\newpage 
\baselineskip=16pt 
\bigskip 
\bigskip 
{\bf References} 
\medskip 

\medskip\par\pp{[1]} 
S. W. Hawking, Comm. Math. Phys. {\bf 43}, 199 (1974); 
S. W. Hawking, Nature {\bf 248}, 30 (1974). 

\medskip\par\pp{[2]} 
J. Kormendy and D. Richstone, Ann. Rev. Astron. Astrophys. {\bf 33}, 581 
(1995); D. Richstone et al., Nature {\bf 395}, A14 (1998); R. Narayan, 
D. Barret, and J. E. McClintock, Astrophys. J. {\bf 482}, 448 (1997).

\medskip\par\pp{[3]} 
R. Bousso and S. W. Hawking, Phys. Rev. D {\bf 52}, 5659 (1995); 
R. Bousso and S. W. Hawking, Phys. Rev. D {\bf 54}, 6312 (1996); 
E. W. Kolb and R. L. Slansky, Phys. Lett. {\bf 135 B}, 378 (1984).  

\medskip\par\pp{[4]}
A. Strominger and C. Vafa, Phys. Lett. {\bf 379 B}, 99 (1996). 

\medskip\par\pp{[5]} 
D. E. Osterbrock, {\sl Astrophysics of Gaseous Nebulae and Active 
Galactic Nuclei} (Univ. Science Books, Mill Valley, 1989). 

\medskip\par\pp{[6]} 
F. C. Adams and G. Laughlin, Rev. Mod. Phys. {\bf 69}, 337 (1997). 

\medskip\par\pp{[7]} 
P. Langacker, Phys. Rep. {\bf 72}, 186 (1984); 
D. Perkins, Ann. Rev. Nucl. Parti. Sci. {\bf 34}, 1 (1984). 

\medskip\par\pp{[8]} 
Ya. B. Zeldovich, Phys. Lett. {\bf 59 A}, 254 (1976);  
Ya. B. Zeldovich, Sov. Phys. JETP, {\bf 45}, 9 (1977); 
S. W. Hawking, D. N. Page, and C. N. Pope, Phys. Lett. {\bf 86 B}, 
175 (1979); D. N. Page, Phys. Lett. {\bf 95 B}, 244 (1980); 
F. C. Adams, G. Laughlin, M. Mbonye, and M. J. Perry, 
Phys. Rev. D {\bf 58}, 083003 (1998). 

\medskip\par\pp{[9]} 
N. D. Birrell and P.C.W. Davies, {\sl Quantum Fields in Curved Space} 
(Cambridge Univ. Press, Cambridge, 1982); 
K. S. Thorne, R. H. Price, and D. A. MacDonald, {\sl Black Holes: 
The Membrane Paradigm} (Yale Univ. Press, New Haven, 1986); 
D. N. Page, Phys. Rev. D {\bf 13}, 198 (1976). 

\medskip\par\pp{[10]} 
F. J. Dyson, Rev. Mod. Phys. {\bf 51}, 447 (1979); 
J. Binney and S. Tremaine, {\sl Galactic Dynamics} 
(Princeton Univ. Press, Princeton, 1987). 
 
\medskip\par\pp{[11]} 
E. W. Kolb and M. S. Turner, {\sl The Early Universe} 
(Addison-Wesley, Redwood City, 1990). 

\medskip\par\pp{[12]} 
D. N. Page and M. R. McKee, Phys. Rev. D {\bf 24}, 1458 (1981); 
D. N. Page and M. R. McKee, Nature {\bf 291}, 44 (1981). 

\end{document}